\documentclass[12pt]{article}
\usepackage[utf8]{inputenc}
\usepackage[english]{babel}
\usepackage{amsmath}
\usepackage{amsfonts}
\usepackage{amssymb}
\usepackage{color}
\usepackage{graphicx}
\usepackage{subfigure}
\usepackage{authblk}
\usepackage[numbers,sort&compress]{natbib}

\title{Cross-validation tests for cryo-EM maps using an independent particle set}

\author[1]{Sebastian Ortiz}
\author[2]{Luka Stanisic}
\author[3]{Boris A Rodriguez}
\author[2]{Markus Rampp}
\author[4,5]{Gerhard Hummer}
\author[1,4,*]{Pilar Cossio}

\affil[1]{ \normalsize {\it Biophysics of Tropical Diseases, Max Planck Tandem Group, Universidad de Antioquia UdeA, Calle 70 No. 52-21, Medellín, Colombia.}}
\affil[2]{\normalsize \textit{Max Planck Computing and Data Facility, 85748 Garching, Germany.} }
\affil[3]{ \normalsize {\it Grupo de Física Atómica y Molecular, Instituto de Física, Facultad de Ciencias Exactas y Naturales, Universidad de Antioquia UdeA, Calle 70 No. 52-21, Medellín, Colombia.}}
\affil[4]{\normalsize {\it Department of Theoretical Biophysics, Max Planck Institute of Biophysics, 60438 Frankfurt am Main, Germany.}}
\affil[5]{\normalsize \textit{Institute of Biophysics,
Goethe University, 60438 Frankfurt am Main, Germany.}}
\affil[*]{{\small\textit{email:} pilar.cossio@biophys.mpg.de; grupotandem.biotd@udea.edu.co}}

\date{}

\begin{document}
\maketitle
\newpage

\begin{abstract}
Cryo-electron microscopy is a revolutionary technique that can provide 3D density maps at near-atomic resolution. However, map validation is still an open issue in the field. Despite several efforts from the community, it is possible to overfit the reconstructions to noisy data.
Here, inspired by modern statistics, we develop a novel methodology that uses a small independent particle set to validate the 3D maps. The main idea is to monitor how the map probability evolves over the control set during the refinement.
The method is complementary to the gold-standard procedure, which generates two reconstructions at each iteration. We low-pass filter the two reconstructions for different frequency cutoffs, and we calculate the probability of each filtered map given the control set. For high-quality maps, the probability should increase as a function of the frequency cutoff and of the refinement iteration. We also compute the similarity between the probability distributions of the two reconstructions. As higher frequencies are added to the maps, more dissimilar are the distributions. We optimized the BioEM software package to perform these calculations, and tested the method on several systems, some which were overfitted. Our results show that our method is able to discriminate the overfitted sets from the non-overfitted ones. 
We conclude that having a control particle set, not used for the refinement, is essential for cross-validating cryo-EM maps. 
\end{abstract}
\newpage

Cryo-electron microscopy (cryo-EM) has revolutionized structural biology by providing electron density maps of biomolecules  that were difficult to resolve with X-ray crystallography or nuclear magnetic resonance \cite{Kuhlbrandt2014,Cheng2015, murata2018review}. The introduction of direct electron detection cameras \cite{Wu2016,McMullan2016} and novel computational algorithms \cite{Kervrann2016,Cossio2018} has enabled the reconstruction of density maps with near-atomic details. To date, thousands of maps, and their corresponding atomic models, have been deposited in the electron microscopy \cite{Lawson:2011} and protein data banks \cite{Berman2000} (EMDB and PDB, respectively). 
 

Typically, cryo-EM maps are reconstructed using the gold-standard procedure \cite{henderson2012outcome, scheres2012prevention}. The particle images are divided into two sets, and two independent reconstructions are generated. The reconstructions are refined iteratively using maximum-likelihood \cite{Sorzano2004,Tang2007} or Bayesian techniques \cite{scheres2012relion,Punjani2017}. At each iteration the Fourier Shell Correlation (FSC) \cite{Saxton1982, Harauz1986}  between the two independent reconstructions is computed. Fixed FSC threshold criteria at 0.143 \cite{Rosenthal2015} or 0.5 \cite{Harauz1986} are used to determine the resolution of the reconstructions (\textit{i.e.}, the size of the smallest reliable detail). The refinement process is halted when the resolution of the reconstructions stops improving. In the end, the maps are masked and a final resolution is determined.  


However, in spite of several efforts from the cryo-EM community, map validation is still problematic. In the recent Map Challenge it has been shown that there is no absolute ‘gold standard’ \cite{heymann2018}. The protocols are user-dependent and there can be biases due to processing workflows. 
For instance, in the FSC calculation, the resolution estimate is dependent on the radius of the shell in Fourier space, and on the point symmetry of the molecule \cite{van2005fourier, sorzano2017review}. The use of a fixed threshold for the FSC is restricted by the assumption that the noise and the signal are orthogonal \cite{van2005fourier}. In addition, the mask can be a source for overestimating the resolution \cite{Penczek2010, Pintilie2016, Rosenthal2015}. Therefore, the best criteria to estimate the map resolution are still debated in the cryo-EM community \cite{sorzano2017review,van2005fourier}. These issues can lead to overfitted cryo-EM reconstructions. For example, the reported values of the resolution in the model (from the PDB) and in the map (from the EMDB) are different for about 30$\%$ of the deposited data \cite{Afonine2018}.  Moreover, it has been found that more than 70$\%$ of the maps in the EMDB have moderate to low agreement with the model, mostly because of the limited resolvable features of the maps \cite{Neumann2018}. In extreme cases, maps can be reconstructed from pure-noise images \cite{henderson2013overfitting,Shatsky2009}.

Therefore, methods that validate the quality of the maps and models are fundamental for cryo-EM. Randomization of the phases beyond a frequency threshold can give signatures of overfitting in the FSC curve \cite{scheres2012prevention, Chen2013}. Better resolution estimates are obtained with reference-free pipelines using the 1/2 bit non-fixed FSC threshold \cite{van2005fourier,Afanasyev2017}. The local resolution in a map can be evaluated using the background noise of the reconstruction \cite{Kucukelbir2014} or by masking different regions with the FSC \cite{Cardone2013,Pintilie2016}. Predictability of the particle alignment provides quality indicators of the reconstruction \cite{Vargas2017,Vargas2016}. Moreover, several metrics that monitor cross-correlations in real or Fourier space between the maps and models indicate the reliability of the resolution \cite{Afonine2018, Neumann2018,Brown2015}. Recently, deep learning algorithms have been introduced to automatically classify maps into high, medium, and low resolution \cite{Avramov2019}.
However, all these methods have the limitation that they do not use the raw data, which ultimately comes from the individual particles, but they only use the maps or models that are product of processing and averaging. For instance, in cryo-EM there is no cross-validation method, such as the R-free in X-ray crystallography \cite{Brunger1992}, which uses an independent control set from the pure experimental data.    

Inspired by modern statistical methods, we here propose an unbiased strategy that validates cryo-EM reconstructions using a small control set of particle images that are omitted from the refinement process. We do not focus on determining a specific value for the resolution but we develop a simple cross-validation technique that monitors how the quality of the reconstructions evolves during the refinement procedure.  We first calculate the BioEM \cite{cossio2013bayesian,cossio2017bioemgpu} probability of the maps, given the control set, as a function of a low-pass frequency cutoff of the reconstructions. High-quality maps should increase in probability for higher frequency cutoffs and higher refinement iterations. We then show that the similarity between the probability distributions of the two reconstructions from the gold-standard procedure is an additional quality indicator. Finally, we test the method on different systems and asses its effectiveness to discriminate overfitted maps.

\section*{Results}  

\subsection*{Cross-validation protocol.}
We propose a statistical framework for the cross-validation of cryo-EM reconstructions. First, and foremost, the validation analysis is done over a small control set of particle images not used in the refinement process.  Analogously to the R-free in X-ray crystallography \cite{Brunger1992}, this independent set should give an unbiased estimate of the quality of the reconstructions. 

\begin{figure}[h!]
\includegraphics[width=\columnwidth]{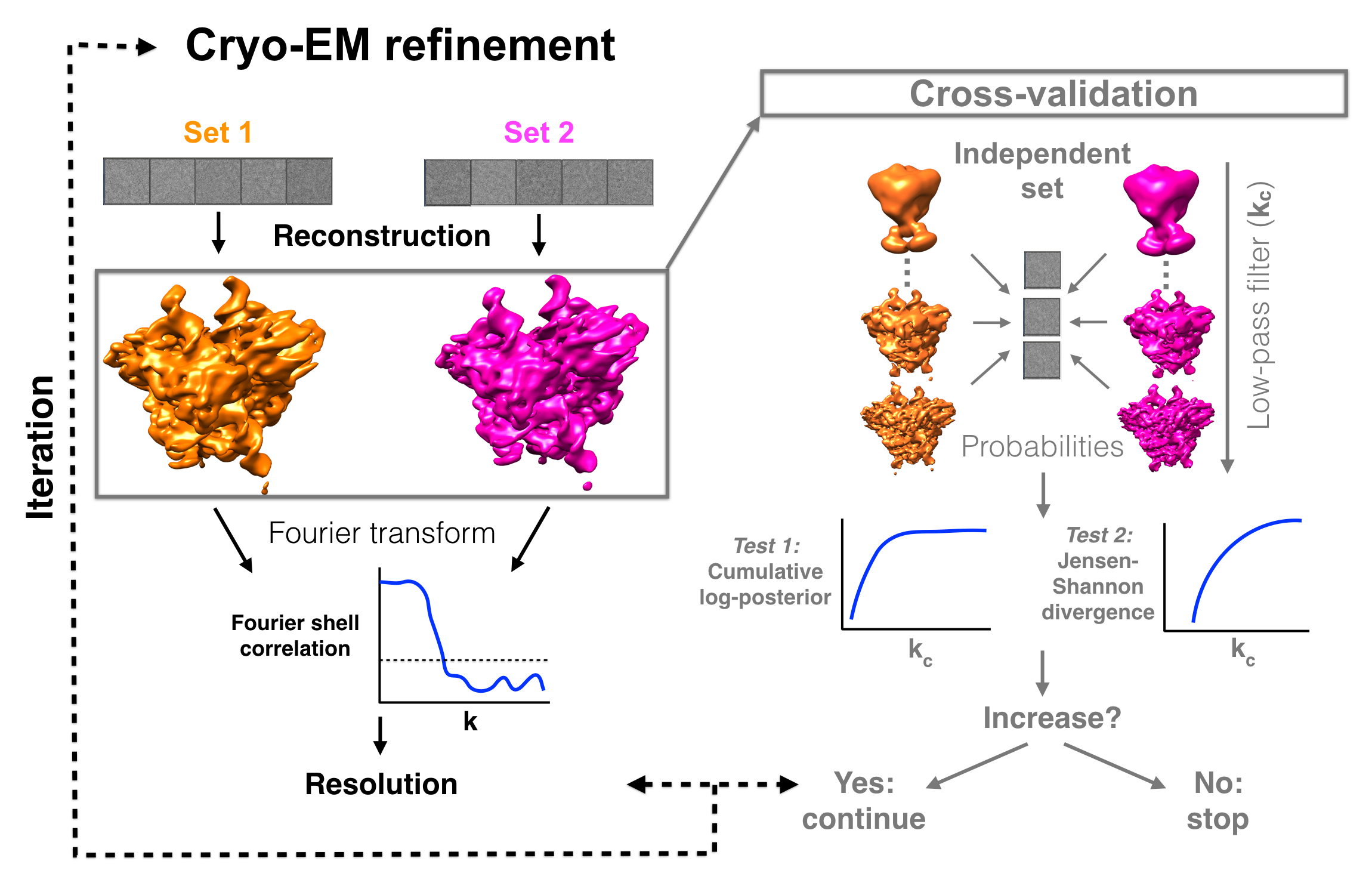}
\caption{ Cross-validation protocol for unbiased map validation in cryo-EM. (\textbf{left}) Gold-standard refinement procedure in cryo-EM.  Two particle sets are used to generate two independent reconstructions. These reconstructions are compared using the Fourier shell correlation (FSC). A fixed FSC threshold is used to extract the resolution of the reconstructions. The process is iterated until the resolution stops improving. (\textbf{right}) Novel cross-validation protocol using a small control particle set. At each iteration of the refinement, the reconstructions are low-pass filtered to different frequency cutoffs $k_c$. The BioEM probabilities \cite{cossio2013bayesian,cossio2017bioemgpu}, over the independent control set, are calculated as a function of $k_c$. Two tests validate the quality of the reconstructions: 1) the cumulative log-posterior and 2) the statistical similarity between the probability distributions (measured with a normalized Jensen-Shannon divergence). The results from both tests should increase as a function of the frequency cutoff. The maps represented correspond to the RAG1-RAG2 complex (see the Methods).}
\label{fig:cross-validation}
\end{figure}

Fig. \ref{fig:cross-validation} shows the work-flow of the methodology. The refinement  is done following the gold-standard procedure (Fig. \ref{fig:cross-validation}--left), where two reconstructions are generated at each iteration step. These two reconstructions are validated using the control particle set (Fig. \ref{fig:cross-validation}--right). At each iteration, the two maps are low-pass filtered to different frequency cutoffs, $k_c$ (see the Methods). The BioEM \cite{cossio2013bayesian} probability, $P_{i\omega}(k_c)$, for each set $i=1,2$  is calculated over the control set, $\omega \in \Omega$, with $N_\omega$ particles. As a first cross-validation test, we monitor the cumulative log-posterior, $\sum_\omega \ln (P_{i\omega}(k_c))/N_\omega $, as a function of $k_c$ for each set $i$. This cumulative evidence should increase or remain constant as higher frequencies are added to the maps. Failing this test is a prime indicative that there is a problem in the refinement process. 

The second cross-validation test consists on measuring the similarity between the probability distributions of the two reconstructions, also as a function of the frequency cutoff. For this purpose, we calculate a normalized Jensen-Shannon divergence ($\mathrm{NJSD}$) (see the Methods). The NJSD is a positive, symmetric and bound metric that measures how distinguishable are the probability distributions from the reconstructions sets $1$ and $2$. We expect that as more frequencies are added to the reconstructions, more noise is added, and the probability distributions are more uncorrelated (\textit{i.e.}, less similar). 

In the following, we describe in detail the two cross-validation tests.

\subsection*{Map evidence from the cumulative log-posterior.}
We tested the methodology over several cryo-EM datasets: the synaptic RAG1-RAG2 complex (RAG1-RAG2) \cite{heng2015RAG}, the human HCN1 channel (HCN1) \cite{lee2017HCN1}, and the TRPV1 ion channel (TRPV1) \cite{liao2013system3}. These systems represent a diverse set of biomolecular families, with membrane proteins and protein-nucleicacids complexes. The reconstruction refinement was performed using the gold-standard procedure in RELION \cite{scheres2012relion}.
The final resolution of these systems ranges from approximately $3$ to $6$ \AA (see the Methods). To analyze the impact of overfitting, we studied two additional systems: cryo-EM reconstructions from the HIV-1 envelop trimer (HIV-ET) \cite{Mao:hiv} and a set of synthetic pure-noise images that act as a `false' control set with the RAG1-RAG2 reconstructions (see the Methods). This was motivated by the fact that some reconstructions might have been generated from pure-noise particles, and their resolution might have been over-estimated  \cite{Subramanian2013,henderson2013overfitting,Shatsky2009}. 

\begin{figure}[h!]
\includegraphics[width=\columnwidth]{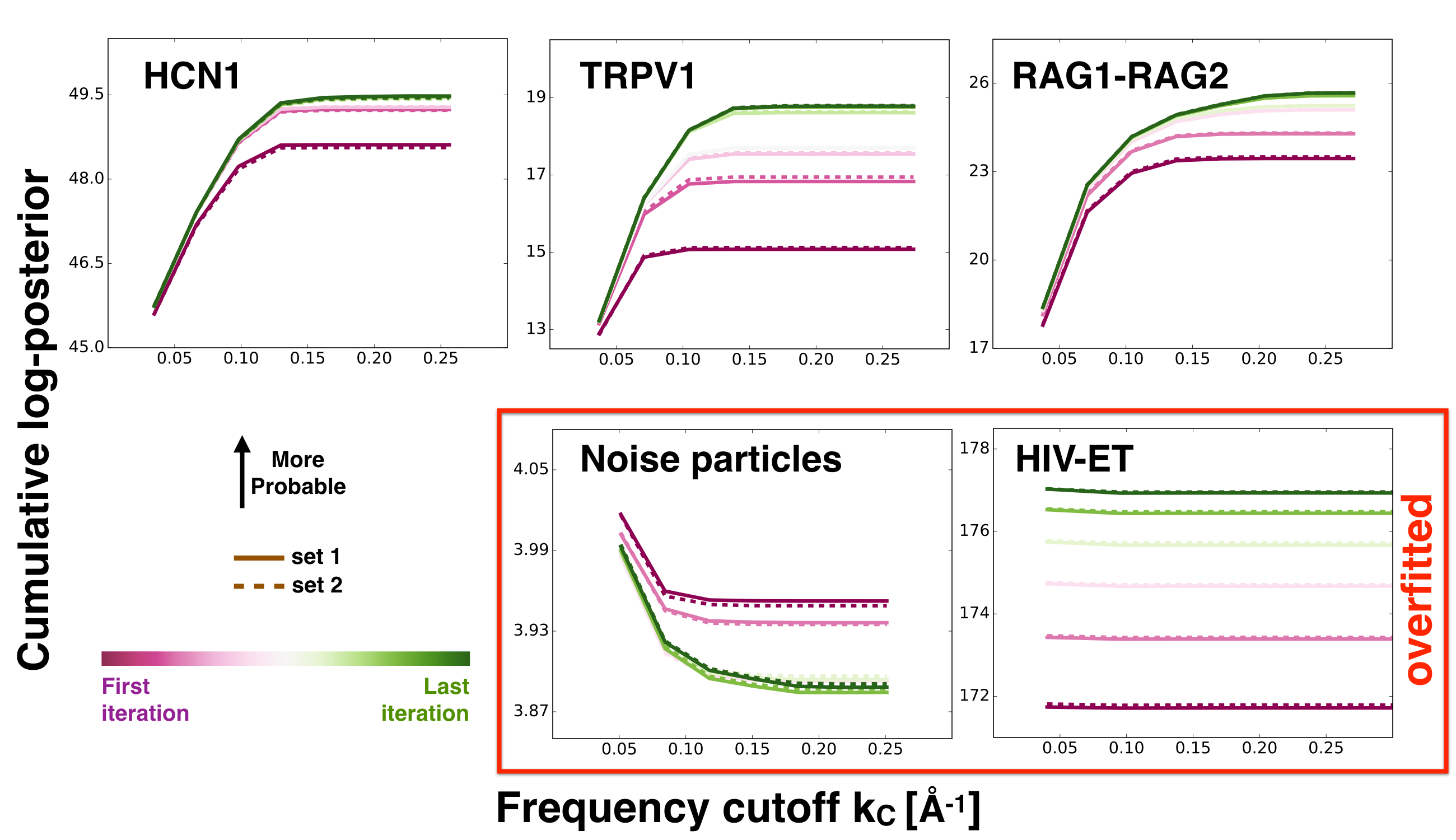}
\caption{ The cumulative log-posterior relative to noise $\sum_\omega \ln (P_{i\omega})/N_\omega - \ln (P_{\mathrm{Noise}}) $, over the control set with $N_\omega$ images, as a function of the frequency cutoff for reconstructions from set $i=1$ and $2$ (solid and dashed lines, respectively). The results are shown for different refinement iteration steps with a gradient color code: the first iteration is maroon and the last iteration is green. On the top row, we show the results for the standard cryo-EM systems: HCN1, TRPV1 and RAG1-RAG2 for $N_\omega=5000$. Systems that exhibit signs of overfitting, {\it i.e.} a noise-particle control set with $N_\omega=1000$ and HIV-ET with $N_\omega=5000$, are shown in the bottom row, highlighted with a red box.}
\label{fig:cumP}
\end{figure}

In Fig. \ref{fig:cumP}, we examine the improvement of the maps by monitoring the cumulative log-posterior relative to noise, $\sum_\omega \ln(P_{i\omega}(k_c))/N_\omega - \ln (P_{\mathrm{Noise}})$, over the control set with $N_\omega=5000$, as a function of $k_c$ for the reconstructions from sets $i=1,2$. The results are shown for different refinement iterations with a gradient color scheme (first iteration: maroon; last iteration: green). These results measure how probable each filtered map is relative to $P_{\mathrm{Noise}}$ (see the Methods).
For the RAG1-RAG2, HCN1 and TRPV1 systems, we find an increase of the map evidence (given by the cumulative log-posterior) as a function of the frequency cutoff. For very high frequencies, the cumulative evidence plateaus. We only observe minor differences between the results from set $i=1$ and $2$ (solid and dashed lines, respectively, in Fig. \ref{fig:cumP}). This is an indication of the similarity between the reconstructions generated from the two sets. Importantly, the results highlight the ability of the BioEM posterior to correctly rank maps of different resolutions. The reconstructions from the last iterations (\textit{i.e.}, the most refined) are the most probable. This is in agreement with what one expects from the 3D-refinement algorithms \cite{Cossio2018}. 

In contrast, for the HIV-ET and noise-particle set, we find a different behavior of the map evidence. We find that the cumulative log-posterior does not increase as a function of the frequency cutoff but decreases or remains constant. For the noise-particle set, the map evidence relative to $P_{\mathrm{Noise}}$ is small, and the differences between iterations are almost two orders of magnitude smaller than for the non-overfitted sets. Moreover, for this case, as the refinement iterations increase, the maps are slightly less probable. This analysis monitors overfitting in cryo-EM: if the map evidence does not increase as a function of the frequency cutoff or the refinement iteration, then there are signs of overfitting in the data.   

\subsection*{Similarity between the probability distributions.}
As a second validation test, we compare the distributions of the posterior probabilities generated by the reconstructions from sets $i=1,2$ over the control set. In the Supplementary Information, we show an example of the probability distributions for the HCN1 system for two frequency cutoffs at a given iteration (Supplementary Fig. 1-top). We find that the probability distributions, over the independent set, are quite similar for both reconstructions. However, there are small differences between them, and the higher-frequency maps present larger fluctuations (Supplementary Fig. 1-bottom). These differences can be quantified using a normalized Jensen-Shannon divergence (NJSD; see the Methods).

\begin{figure}[h!]
\includegraphics[width=\columnwidth]{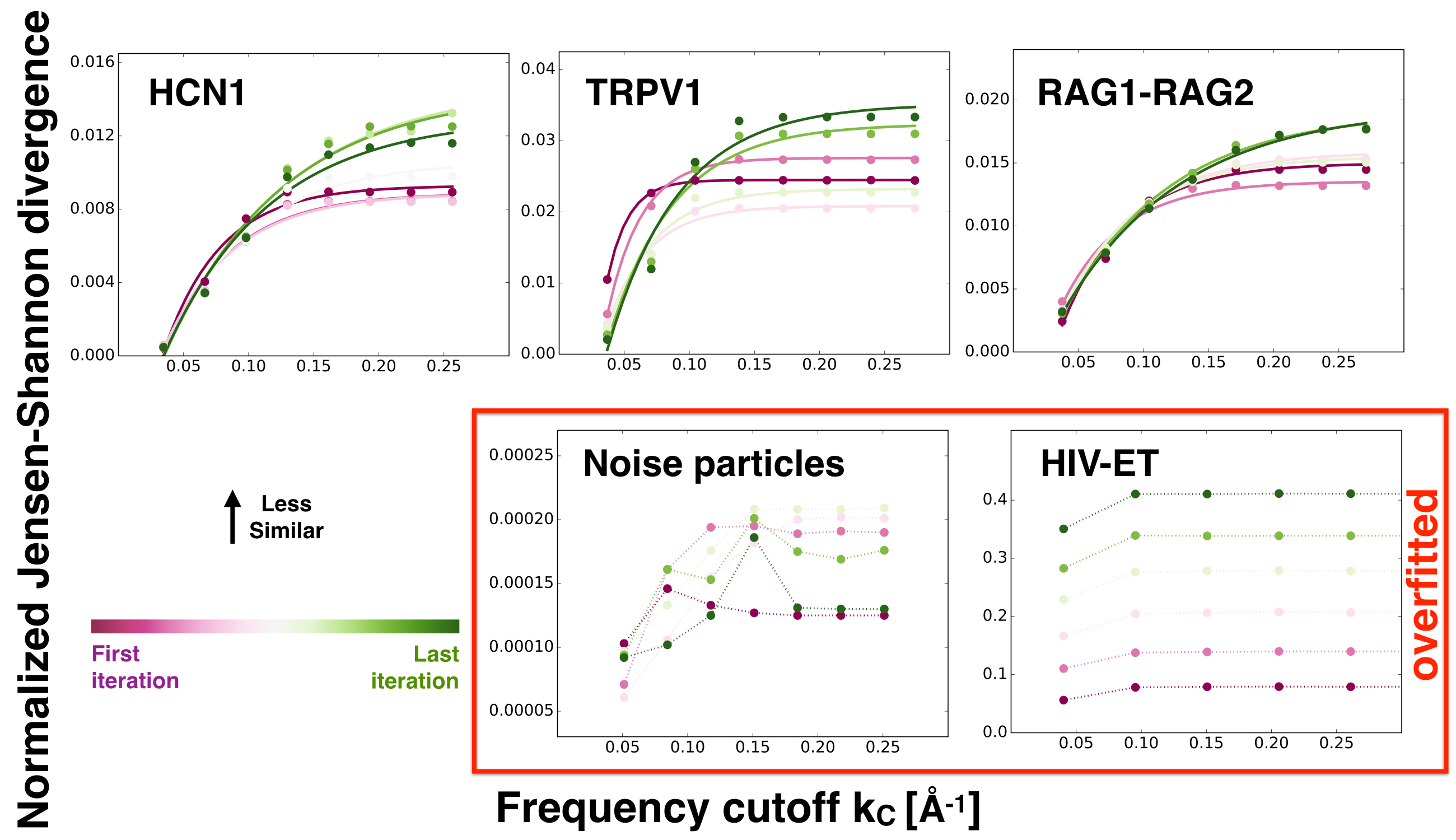}
\caption{Normalized Jensen-Shannon divergence (NJSD) as a function of the frequency cutoff. This metric calculates the similarity between the distributions of the BioEM probabilities computed for the two reconstructions from sets 1 and 2.  We use a gradient color code for the refinement iteration steps: the first iteration is maroon and the last iteration is green. On the top row, we show the results for the standard cryo-EM systems: HCN1, TRPV1 and RAG1-RAG2. For these systems, we fit the data points to an inverse exponential function $-Ae^{-k_c/\gamma}+B$ (solid lines).
Systems that present signs of overfitting, a noise-particle control set and HIV-ET, are shown in the bottom row with dashed lines as a guide. The red box highlights the overfitted systems. The number of images in the control sets are the same as for the data in Fig. \ref{fig:cumP}.}
\label{fig:NJSD}
\end{figure}

In Fig. \ref{fig:NJSD}, we plot the $\mathrm{NJSD}$ as a function of the frequency cutoff $k_c$. Interestingly, for the RAG1-RAG2, HCN1 and TRPV1 systems, we observe that as the filtered maps contain higher frequencies, the larger the value of the $\mathrm{NJSD}$. This implies that the probability distributions between maps with higher frequencies are less similar, possibly because they are more uncorrelated due to the high-frequency noise. For these standard systems, we also find that as the iteration increases the NJSD reaches at higher frequencies a plateau value. This behavior can be fit with an inverse exponential function $-Ae^{-k_c/\gamma}+B$ (see below and solid lines in Fig. \ref{fig:NJSD}).
On the contrary, for the HIV-ET and noise-particle set, we find that the NJSD remains constant or has random behavior, suggesting that distributions do not consistently change when higher frequencies are added to the maps.

\subsection*{Cross-validation tests versus resolution.}
\begin{figure}[h!]
\centering
\includegraphics[width=0.8\columnwidth]{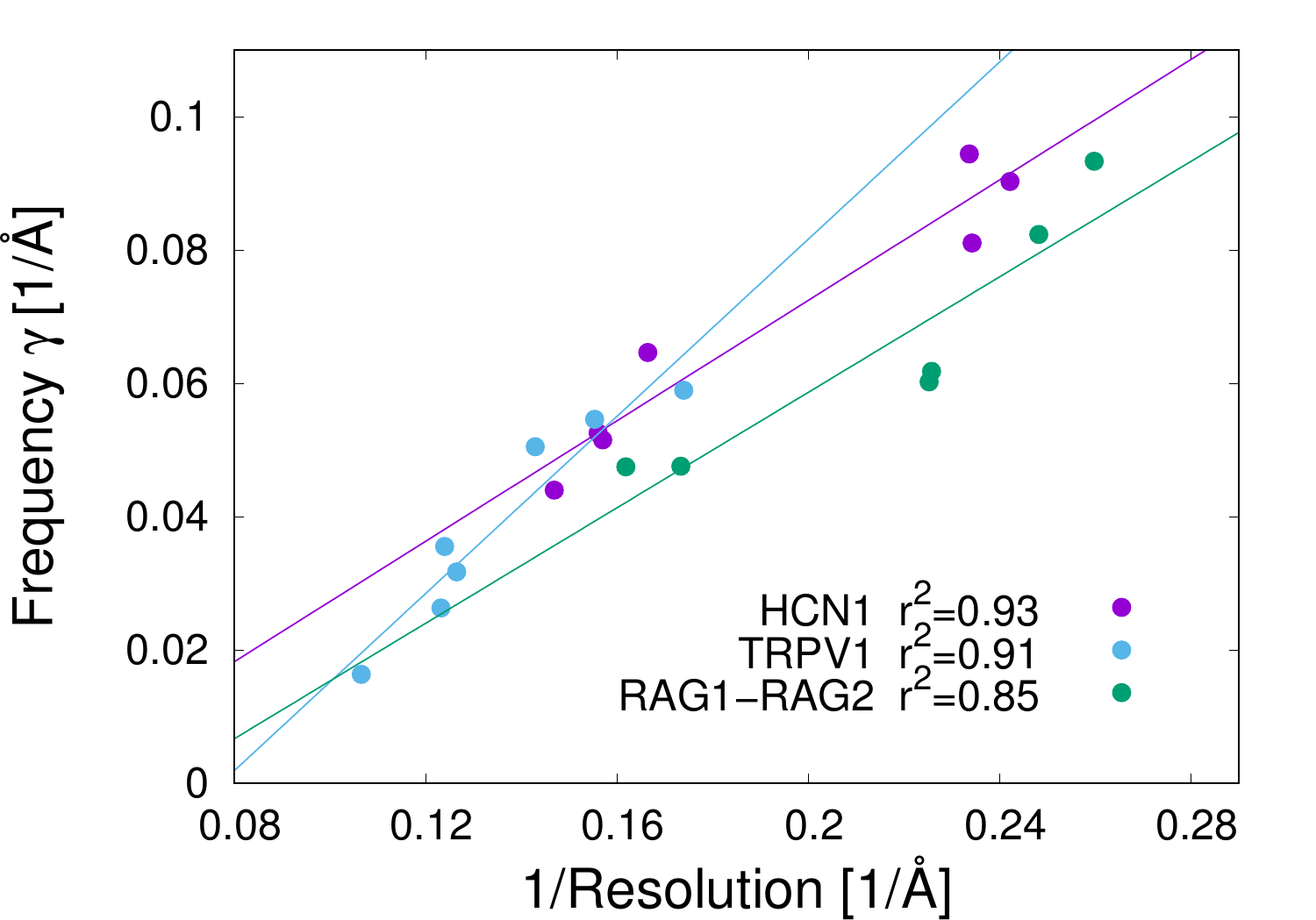}
\caption{Frequency $\gamma$ versus the inverse of the resolution for the standard cryo-EM systems: HCN1, TRPV1 and RAG1-RAG2. The NJSD curves for these systems were fitted to an inverse exponential function $-Ae^{-k_c/\gamma}+B$. We find large correlations between $\gamma$ and the inverse of the resolution (calculated using the 0.143 criteria). The correlation coefficients are  $r^2=0.93$, $0.91$, and $0.85$, for HCN1, TRPV1 and RAG1-RAG2, respectively. Solid lines show the linear fits.}
\label{fig:Cor-res-b}
\end{figure}

 We explored how the cross-validation results depend on the map resolution. For the HCN1, TRPV1 and RAG1-RAG2 systems, we find that the $\mathrm{NJSD}$ curves can be fitted to an inverse exponential function, $-Ae^{-k_c/\gamma}+B$ (solid lines shown in Fig. \ref{fig:NJSD}). Intuitively, the frequency $\gamma$ indicates where the plateau of the NJSD is reached. In Fig. \ref{fig:NJSD}, we can qualitatively see that $\gamma$ is larger for higher refinement iterations.  
In  Fig. \ref{fig:Cor-res-b}, we plot the frequency $\gamma$ as a function of the inverse of the resolution (calculated using the FSC at the threshold 0.143). 
Interestingly, we find that the frequency $\gamma$ is highly correlated to the inverse of the resolution with correlation coefficient $r^2=0.93$, $0.91$, and $0.85$, for  HCN1, TRPV1 and RAG1-RAG2, respectively. 
These results show that even from a small independent control set, it is possible to extract unbiased information about the map resolution. 
We note that for the HIV-ET and noise-particle sets it is not possible to fit the NJSD data to an inverse exponential function. Therefore, we can only estimate the correlation between $\gamma$ and the inverse of the resolution for the standard cryo-EM systems.  

\subsection*{Convergence over a small cross-validation set.}
We assessed how the results depend on the number of particles in the control set. In Supplementary Fig. 2, we show an example of the cumulative log-posterior and NJSD as a function of the number of images in the control set. We find that after approximately 1000 particles these observables converge, suggesting that only a small set is needed to perform the cross-validation analysis. 
This is confirmed in Supplementary Fig. 3, where we plot the cumulative log-posterior and NJSD as a function of the frequency cutoff for a validation set of 1000 images. For the same set, in Supplementary Fig. 4, we plot the frequency $\gamma$ as a function of the inverse of the map resolution, showing high correlations for the standard cryo-EM systems. These results are very similar to those obtained for the cross-validation set with 5000 particles. 

\section*{Discussion}
In this work, we have developed a novel methodology for cross-validating cryo-EM reconstructions. Importantly, the procedure is performed over an independent particle set that is not used to generate the reconstructions. Two cross-validation tests are proposed. The first consists of monitoring the cumulative log-posterior of the maps as a function of a low-pass filter frequency cutoff. The posterior should increase as a function of the frequency cutoff and the refinement iteration. In the second test, we assess the similarity between the probability distributions generated from the two reconstructions from the gold-standard procedure. The distributions should become less similar as higher frequencies are added to the reconstructions. 
 
We performed the cross-validation tests over several systems: three standard cryo-EM reconstruction sets, and two datasets with noise particles that mimic overfitting. The results show substantial differences. While for the standard cryo-EM sets the results are as expected, the overfitted sets present almost no increment (even sometimes decrease) of the cumulative posterior or the NJSD. Thus, signatures of overfitting can be monitored with the proposed cross-validation tests.  

Our methodology is general and robust. The mathematical framework is not only valid for the BioEM posterior but also for any posterior probability that measures the likelihood of a 3D density given a particle set. The tests converge over a small particle set, typically only 1000 particles. Moreover, the methodology has the potential to be applicable for directly refining atomic models (instead of 3D maps) using an independent control set.

Determining an unbiased estimate of the reconstruction resolution remains an open issue. However, our procedure could shed light on how to tackle  this problem with a different perspective. For example, the resolution could be defined as a multiple of $\gamma$ that determines the frequency at which the information between the probability distributions is governed by noise.  

All-in-all, our work provides a novel way to monitor overfitting in cryo-EM. We conclude that having a control particle set which is not used to generate the reconstructions should become a standard for any cryo-EM application.

\section*{Methods}
{\footnotesize

\subsection*{Benchmark systems.}

We used the following benchmarks that represent diverse biomolecular families and cryo-EM systems:

 \textit{The human hyperpolarization-activated cyclic nucleotide-gated channel} (HCN1) is  a voltage-dependent  ion channel, which was resolved to high resolution using cryo-EM \cite{lee2017HCN1}. The system was resolved in two conformational states, an \textit{apo} state and a cAMP-bound state, to $\sim 3.5$ \AA $\,$ using RELION 3D-refinement \cite{scheres2012relion}. 55870 particles images belonging to the \textit{apo} state together with the defocus information of each particle are available in the Electron Microscopy Public Image Archive (EMPIAR) \cite{Iudin2016} with code 10081. Their pixel and image size are also available in that archive. 

\textit{The recombination-activating genes RAG1-RAG2} form a complex (RAG1-RAG2) that plays an essential role in the generation of antibodies and antigen-receptor genes in a process called V(D)J recombination. Two main structures of the RAG1-RAG2 complex can be distinguished during the V(D)J recombination, a synaptic paired complex and the signal end complex (SEC). These states were resolved to 3.7 and  3.4 \AA, respectively, using cryo-EM \cite{heng2015RAG}. 81946 processed picked particles from the SEC state are deposited in the EMPIAR data bank with code 10049. The defocus information is available for these particles. 

\textit{The  mammalian transient receptor potential TRPV1}  ion channel (TRPV1) is the receptor for capsaicin. Its structure was determined to 3.4 \AA $\,$ using cryo-EM \cite{liao2013system3}. A set of 35645 processed particles for this system are found in the EMPIAR data bank with code 10005. The defocus information is also available for these particles. 

\textit{The human immunodeficiency virus type 1 envelope glycoprotein trimer} (HIV-ET) is a membrane-fusing machine which mediates virus entry into host cells. The structure of the apo HIV-1 envelope glycoprotein in the trimer-conformation was determined to 6 \AA $\,$ using the 0.5 FSC threshold with cryo-EM \cite{Mao:hiv}. A set of 124478 particles used in the refinement process is available in EMPIAR with code 10008. The defocus information is also available for these particles. 

For all of the above cases, a subset of 5000 particles was randomly selected to be used as the cross-validation set. Specifically, these particles are not used in the refinement processes. 

\textit{Pure-noise images:} we generated a set of synthetic 1000 pure-noise particles. Each particle contains random intensities following a Gaussian distribution with zero mean and unit variance (for details see the Supplementary Information). These images were used as a ``false" control set to assess the RAG1-RAG2 reconstructions.

\subsection*{3D refinement.} 
\begin{table}[ht!]
    \centering
    \begin{tabular}{|c|c|c|c|c|}
    \hline
         System & \#Particles & Symmetry & \#iterations & Final resolution*  \\ \hline
         HCN1 & 50870 & C4 &  17 & 4.2\AA \\ \hline
         RAG1-RAG2 & 79946 & C2 & 26 & 3.8\AA \\ \hline
         TRPV1 & 30645 & C4 & 24 & 5.3\AA \\ \hline
         HIV-ET & 119478& C3& 10 & 9.9\AA \\\hline
    \end{tabular}
    \vspace{.01mm}
    \small{*using the 0.143 FSC threshold}
    \caption{Summary of the results from the 3D-refinement using RELION \cite{scheres2012relion} for the cryo-EM systems.}
    \label{tab:my_label}
\end{table}
The RELION \cite{scheres2012relion} software was used to reconstruct the cryo-EM maps. For all systems, we assume that the deposited particles correspond to the same state. Therefore, the preprocessing steps of 2D or 3D classification are not performed. 
As the initial reference map for the 3D refinement, we use the final map reported by the authors low-pass filtered to 60 \AA. This was done to minimize the risk of overfitting \cite{scheres2012prevention}.
The 3D-refinement procedure implements the gold-standard approach by splitting the data into two random halves (sets $i=1,2$) and performing two independent reconstructions. We note that the number of particles used for these reconstructions was slightly less than those of the original works because the particles from the control set were taken out. In all cases, we used the 
RELION default parameters, and point-group symmetries reported by the authors. Table 1 summarizes the results obtained from the 3D refinement. The resolutions are in accordance with the reported ones, taking into account that the post-processing steps were not performed, and that the control set of particles was excluded from the refinement.

\subsection*{Low-pass filter.}

Consider a  map $m$  generated from an iteration of the 3D refinement. Let $\mathcal{F}_m (\mathbf{k})$ be its 3D-Fourier transform, where $\mathbf{k}$ is the reciprocal vector. We perform a low-pass filter on the map, $\mathcal{F}_m^{k_c}(\mathbf{k})$, up to a frequency cutoff $k_c$. The resulting filtered map is 

\setlength\arraycolsep{2pt}
\begin{equation}
\mathcal{F}_m^{k_c}(\mathbf{k})=\left\{\begin{matrix} \mathcal{F}_m(\mathbf{k}) &&  k \leq k_{c} \\ 0 && \text{otherwise}. \end{matrix} \right.
\end{equation}
We use the code \textit{lowpassmap\_fftw} available from the Rubinstein lab webpage \cite{rublab} to perform this calculation. We then convert the map into real space by applying the inverse Fourier transform of $\mathcal{F}_m^{k_c}(\mathbf{k})$. The real-space filtered map is masked and then used as input for the BioEM computation (see below).

\subsection*{BioEM posterior probabilities.}
The BioEM method \cite{cossio2013bayesian} uses a Bayesian framework to quantify the consistency between an experimental image $\omega$ and a given map $m$ (or model) by calculating a posterior probability $P_{m\omega}$.  BioEM takes into account the relevant physical parameters ($\Theta$) for the image formation: center displacement, normalization, offset, noise, orientation and CTF parameters (defocus, amplitude, and B-factor). $P_{m\omega}$ is calculated by integrating-out all parameters 
\begin{equation}
P_{m\omega} \propto \int L(\omega\vert\Theta,m)p(\Theta)p(m) d\Theta ~,
\label{eq:bioEM}
\end{equation}
where $p(m)$ and $p(\Theta)$ are the prior probabilities of the map and parameters, respectively, and $L(\omega\vert\Theta,m)$ is the likelihood function. We considered the prior probabilities of maps and parameters uniform over the integration intervals. In Eq. \ref{eq:bioEM}, the integrals over the offset, noise and normalization are performed analytically \cite{cossio2013bayesian}, and that over the center displacement is described in ref. \cite{cossio2017bioemgpu}. The integral over the orientations and CTF defocus is done using a double-round algorithm, which is described in the following subsection. 

Similarly as in ref. \cite{cossio2013bayesian}, we define a noise model $P_{\mathrm{Noise}}=(2\pi \lambda ^2 e)^{-N_{\mathrm{pix}}/2}$ where $N_{\mathrm{pix}}$ is the number of pixels and $\lambda$ is the image variance (by default $\lambda=1$). $P_{\mathrm{Noise}}$ is used as a reference to compare the posterior probabilities.

\subsection*{BioEM algorithm.}
To optimize the computations, we divided the BioEM posterior calculation into two rounds. The objective of the first round is to obtain the best orientations for each particle. In this round, an all-orientations to all-particles algorithm is performed \cite{cossio2017bioemgpu}. As the BioEM input map, we used the final reconstruction from the refinement with a broad mask and without low-pass filtering. To sample the orientations, we used 36864 quaternions that sample uniformly orientation space \cite{Yershova2010}. The particles were grouped into sets with similar experimental defocus with $0.4\mu m$ range, and an independent orientation search was performed for each group. In this round, the best 10 orientations for each particle are obtained. An example of the BioEM input for the first round is presented in the Supplementary Information. 

In the second round, a zoom around the best 10 orientations from the first round and experimental defocus is performed for each low-pass filtered reconstruction from the different refinement iterations. The zoom around each best orientation is done using 125 quaternions with approximately 0.01 grid spacing, resulting in 1250 zoomed-orientations for each particle. This procedure is described in detail in ref. \cite{Cossio:Micro:2018}. The defocus of each particle is fixed to its experimental value.
We used 8 filtering-frequencies for each reconstruction; these were distributed uniformly from $1/(p_s\sqrt{N_{\mathrm{pix}}})$ to $1/(3p_s)$ where $p_s$ is the pixel size. All reconstructions were masked using the same broad mask as for round 1.
An example of the BioEM input file for round 2 is presented in the Supplementary Information.

\subsection*{BioEM code.}
The BioEM code has been extended with several optimizations, which drastically increase performance for the second round of calculations. Most importantly, the main data structures and algorithm were modified to allow for a parallel comparison of multiple orientations to a single particle image. Initial reading of the input files has been parallelized, and the overall memory consumption decreased. These code changes lead to more efficient utilization of the computing resources, and hence to a faster calculation of posterior probabilities, especially for the workloads specific to the second round. For more information, we refer the reader to the BioEM user manual: https://readthedocs.org/projects/bioem/.

\subsection*{Normalized Jensen-Shannon divergence.}
Measuring a distance among probability distributions is a common task in statistics. Most distance measures include concepts from information theory, such as the Kullback-Leibler divergence \cite{Kullback1968,Lin1991} or the Shannon entropy \cite{cover2006elements}. In this work, we measure the statistical similarity between the probability distributions from reconstructions from set 1 and set 2 calculated over the control set. We define a metric that is the Jensen-Shannon divergence \cite{cover2006elements,Lin1991} normalized by the individual Shannon entropies
\begin{equation}
\mathrm{NJSD}=\frac{\sum_{\omega}[P_{1\omega}\ln (P_{1\omega}/M_{\omega}) + P_{2\omega}\ln (P_{2\omega}/M_{\omega})]}{2(\sum_{\omega} P_{1\omega}\ln (P_{1\omega})\sum_{\omega} P_{2\omega}\ln (P_{2\omega}))^{1/2}},
\label{njsd_eq}
\end{equation}
where $P_{1\omega}$ and $P_{2\omega}$ are the probabilities of the reconstructions from set 1 and 2, respectively, over image $\omega$, and $M_\omega=(P_{1\omega}+P_{2\omega})/2$. 
For simplicity of notation, we have omitted the dependency of the probabilities on the frequency cutoff $k_c$. To calculate Eq. \ref{njsd_eq}, we normalize the posterior probabilities such that  $P_{1\omega}+P_{2\omega}=1$ for each image $\omega$, frequency cutoff and iteration. 

In Eq. \ref{njsd_eq}, the numerator measures the correlation between the probability distributions, and the Shannon entropies in the denominator play the role of a normalization factor.  Some important properties of the NJSD metric are that it is positive, symmetric and its lower bound is 0 if and only if $P_{1\omega}=P_{2\omega}$ for all particles $\omega$.
}

\normalsize
\section*{Data availability}
The BioEM code is available at
https://github.com/bio-phys/BioEM. 
A tutorial to perform the cross-validation protocol is available at: 
\\https://github.com/bio-phys/BioEM-tutorials.

\section*{Acknowledgements}
The authors thank Dr. Alessandro Laio for insightful discussions, Dr. Jose Maria Carazo for information about the overfitted systems, and Dr. Frank Avila for proof reading. 
S.O. and P.C. were supported by Colciencias, University of Antioquia and Ruta N, Colombia. G.H., and P.C. acknowledge the support of the Max Planck Society.
Some computations were performed on a local server with an NVDIA Titan X GPU. PC gratefully acknowledges the support of NVIDIA Corporation for the donation of this GPU. Other computations were performed at the Max Planck Computing and Data Facility.  

\section*{Author contributions}
G.H. and P.C. conceived the presented idea. S.O., G.H, and P.C. developed the theory. S.O. and P.C.  performed the computations. L.S. and M.R. co-developed and optimized the code. All authors discussed the results and contributed to the final manuscript.

\bibliography{biblio}{}
\bibliographystyle{naturemag}

\end{document}


\maketitle
\newpage

\section*{Supplementary Figures}
\begin{figure}[h!]
\centering
\includegraphics[width=0.5\columnwidth]{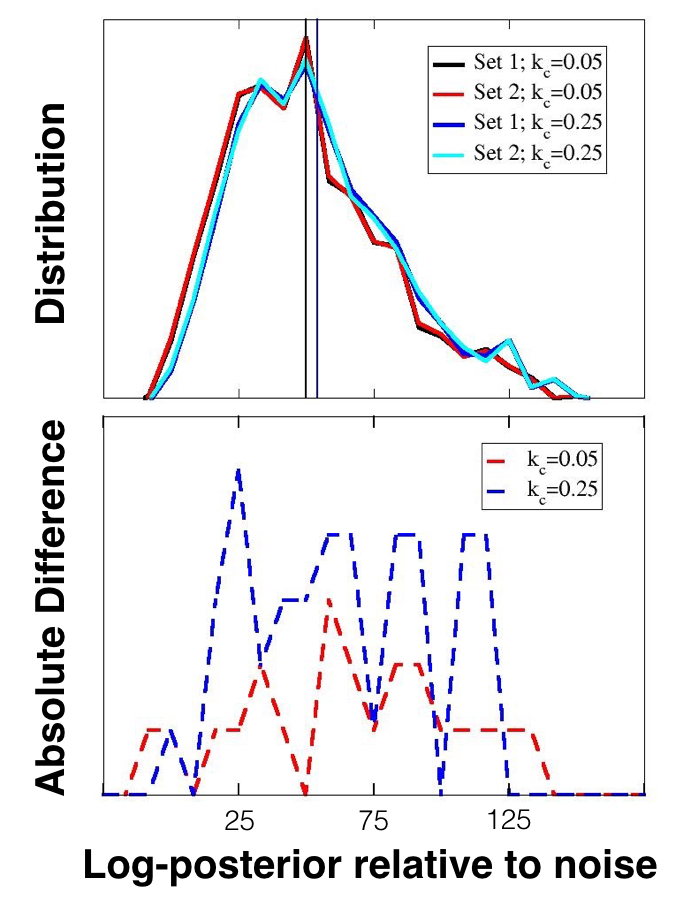}
\caption{\textit{Differences in the log-posterior distributions.}  (\textbf{top})  Examples of the distributions of the log-posterior relative to noise over the independent particle set. The distributions are calculated for the reconstructions from set 1 and set 2 at two cutoff frequencies $k_c=0.05$ and $0.25$ \AA$^{-1}$ for the fifth iteration of refinement of the HCN1 system. The vertical lines are the averages of the distributions. (\textbf{bottom}) Absolute value of the difference between the probability distributions from set 1 and set 2 for $k_c=0.05$ and $0.25$ \AA$^{-1}$. The distributions calculated for the maps with higher frequencies are less similar.}
\label{fig:cross-validation}
\end{figure}

\begin{figure}[h!]
\centering
\includegraphics[width=0.7\columnwidth]{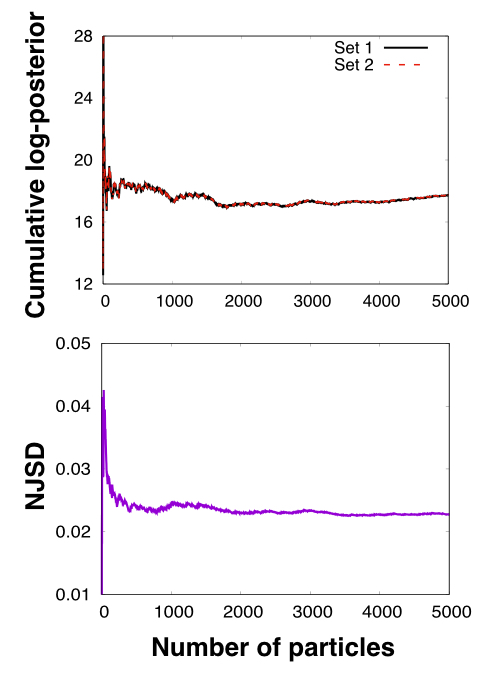}
\caption{\textit{Convergence of the observables.}  (\textbf{top}) The cumulative log-posterior relative to noise $\sum_\omega \ln (P_{i\omega})/N_\omega - \ln (P_{\mathrm{Noise}}) $ for set $i=1$ and $2$ (solid and dashed lines, respectively), and  (\textbf{bottom})
the normalized Jensen-Shannon divergence as a function of the number of particles in the control set. The results are shown for the TRPV1 system for iteration 12 and cutoff frequency $k_c=0.21$ \AA$^{-1}$. The observables converge if more than approximately 1000 particles are used. }
\label{fig:cross-validation}
\end{figure}

\begin{figure}[h!]
\centering
\includegraphics[width=1.\columnwidth]{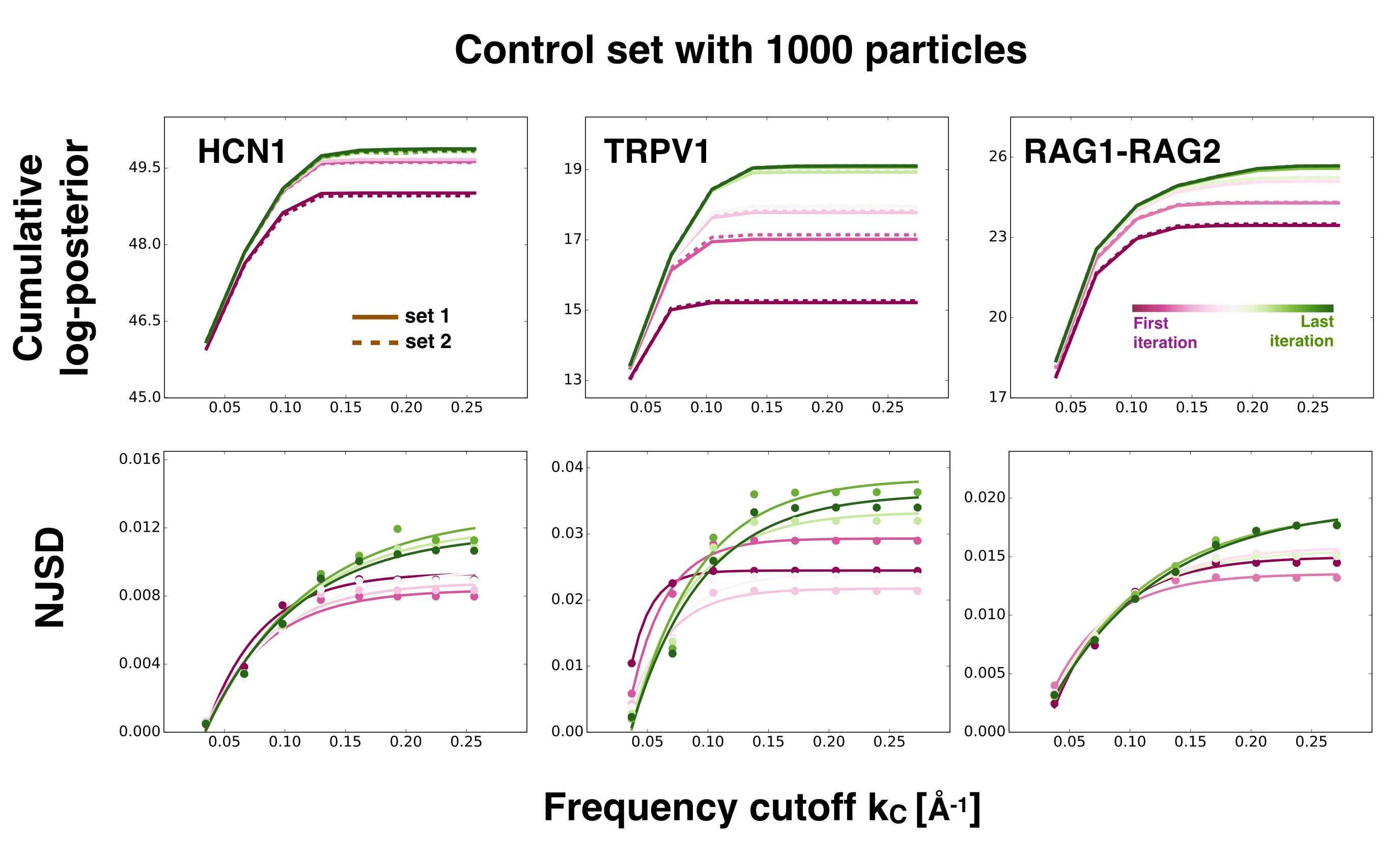}
\caption{\textit{Cumulative log-posterior and NJSD for a control set with 1000 particles.}  (\textbf{top}) The cumulative log-posterior relative to noise  and (\textbf{bottom})
the normalized Jensen-Shannon divergence  as a function of the frequency cutoff.  We use a gradient color code for the refinement iteration steps: the first iteration is maroon and the last iteration is green. The results are shown for the standard cryo-EM systems: HCN1, TRPV1 and RAG1-RAG2. The cumulative log-posterior is shown for the reconstructions from set 1 as solid lines and set 2 as dashed lines. NSJD data is fit to an inverse exponential function $-Ae^{-k_c/\gamma}+B$ (solid lines; bottom). }
\label{fig:cross-validation}
\end{figure}

\begin{figure}[h!]
\centering
\includegraphics[width=\columnwidth]{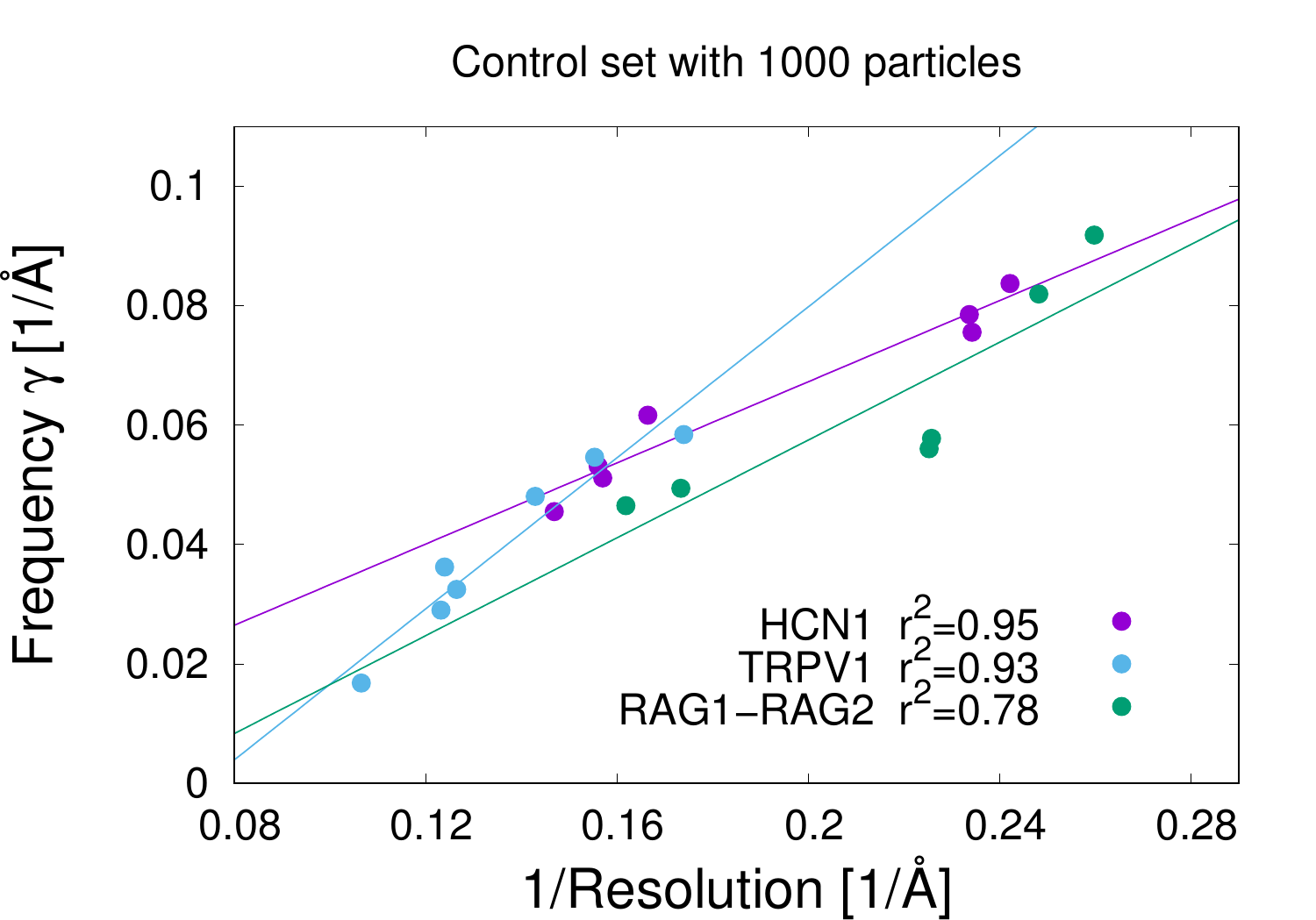}
\caption{\textit{Frequency ($\gamma$) versus the inverse of the resolution for a control set with 1000 particles.} The results are shown for the standard cryo-EM systems: HCN1, TRPV1 and RAG1-RAG2. The correlation coefficients are  $r^2=0.95$, $0.93$, and $0.78$, respectively. Solid lines show the linear fits.  }
\label{fig:cross-validation}
\end{figure}

\newpage
\clearpage
\section*{Supplementary Text}

\subsection*{BioEM input file examples}

\textbf{Round 1:} Example of the BioEM input file for the TRPV1 system for round 1. The best orientations for each particle are obtained using the final map from the refinement. The following input file is for a subset of particles that have experimental defocus between 1.3 and 1.7 $\mu m$. The best 10 orientations for each particle are selected. 
\\ \ \\
\texttt{PIXEL\_SIZE     1.22     \\    
NUMBER\_PIXELS  256        \\ 
USE\_QUATERNIONS             \\
CTF\_DEFOCUS 1.3 1.7 10 \\
CTF\_B\_ENV   0 10 2  \\
CTF\_AMPLITUDE 0.1 0.1 1 \\     
PRIOR\_DEFOCUS\_CENTER 1.5 \\
SIGMA\_PRIOR\_DEFOCUS 0.8 \\
SIGMA\_PRIOR\_B\_CTF     1  \\
DISPLACE\_CENTER  30 1 \\
WRITE\_PROB\_ANGLES 10 \\}

\textbf{Round 2:} Example of the BioEM input file for the TRPV1 system for round 2. The input file is for a single particle that has an experimental defocus of 1.9 $\mu m$.
\\ \ \\
\texttt{PIXEL\_SIZE     1.22     \\    
NUMBER\_PIXELS  256        \\ 
USE\_QUATERNIONS             \\
CTF\_DEFOCUS 1.9 1.9 1 \\
CTF\_B\_ENV   0 10 2  \\
CTF\_AMPLITUDE 0.1 0.1 1 \\     
PRIOR\_DEFOCUS\_CENTER 1.9 \\
SIGMA\_PRIOR\_DEFOCUS 0.3 \\
SIGMA\_PRIOR\_B\_CTF     1  \\
DISPLACE\_CENTER  30 1 }

\subsection*{Pure-noise particles}
We generated a set of 1000 synthetic pure-noise particles.  Each particle has an image size of $180\times180$ and a pixel size of 1.23 \AA. The particles contain random intensities following a Gaussian distribution with zero mean and unit variance. Because there is no experimental defocus, the BioEM probabilities are computed by performing round 1 with defocus range between 0.5 and 4.5 $\mu m$ and using 4608 quaternions uniformly distributed in orientation space. This analysis was performed for each of the refined maps of the RAG1-RAG2 system.